\documentclass[aps,twocolumn]{revtex4}
\usepackage{epsfig}
\usepackage{graphicx}
\usepackage{amsmath}
\usepackage{booktabs}
\usepackage{makecell}
\usepackage{color}

\begin{document}

\title{$T^+_{cc}$ and its partners}

\author{Chengrong Deng$^{a,b}{\footnote{crdeng@swu.edu.cn}}$ and Shi-Lin Zhu$^{b}{\footnote{zhusl@pku.edu.cn}}$}
\affiliation{$^a$School of Physical Science and Technology,
Southwest University, Chongqing 400715, China}
\affiliation{$^b$School of Physics and Center of High Energy
Physics, Peking University, Beijing 100871,China}

\begin{abstract}

Inspired by the $T^+_{cc}$ signal discovered by the LHCb Collaboration, we
systematically investigate the doubly heavy tetraquark states with the molecule
configuration $[Q_1\bar{q}_2]_{\mathbf{1}_c}[Q_3\bar{q}_4]_{\mathbf{1}_c}$ ($Q=c$
and $b$, $q=u$, $d$ and $s$) in a nonrelativistic quark model. The model
involves a color screening confinement potential, meson-exchange interactions
and one-gluon-exchange interactions. The state $T^+_{cc}$ with
$IJ^P=01^+$ is a very loosely bound deuteron-like state with a
binding energy around 0.34 MeV and a huge size of 4.32 fm. Both the
meson exchange force and the coupled channel effect play a pivotal
role. Without the meson exchange force, there does not exist the
$T^+_{cc}$ molecular state. In strong contrast, the QCD valence bond
forms clearly in the $T^-_{bb}$ system when we turn off the
meson-exchange force, which is very similar to the hydrogen molecule
in QED. Moreover, the $T^-_{bb}$ becomes a helium-like QCD-atom if
we increase the bottom quark mass by a factor of three. Especially,
the states $T^-_{bb}$ with $01^+$, $T^0_{bc}$ with $00^+$ and $01^+$
and the $V$-spin antisymmetric states $T^-_{bbs}$ with
$\frac{1}{2}1^+$, $T^0_{bcs}$ with $\frac{1}{2}0^+$ and
$\frac{1}{2}1^+$ can form a compact, hydrogen molecule-like or
deuteron-like bound state with different binding dynamics. The
high-spin states $T^0_{bc}$ with $02^+$ and $T^0_{bcs}$ with
$\frac{1}{2}2^+$ can decay into $D$-wave $\bar{B}D$ and $\bar{B}_sD$
although they are below the thresholds $\bar{B}^*D^*$ and
$\bar{B}^*_sD^*$, respectively. The isospin and $V$-spin symmetric
states are unbound. We also calculate their magnetic moments and
axial charges.
\end{abstract}

 \maketitle

\section{Introduction}

The theoretical explorations on the possible stable doubly heavy
tetraquark states were pioneered in the early
1980s~\cite{potential}. These states were investigated with various
formalisms such as the MIT bag model~\cite{bag}, constituent quark
models~\cite{constituentmodel,constituentmodel1}, chiral perturbation
theory~\cite{chiralperturbation}, string model~\cite{string},
lattice QCD~\cite{lqcdbb}, and QCD sum rule
approach~\cite{qcdsum,qcdsum1}. Although the state $T^-_{bb}$ with
$01^+$ seems stable in various theoretical frameworks, its
production turns out to be very challenging. The discovery of the
doubly charmed baryon $\Xi^{++}_{cc}$ by the LHCb Collaboration
\cite{lhcbxicc} has stimulated the enthusiasms on the doubly heavy
tetraquark states~\cite{ours,karliner,Eichten,Francis,bicudo,unstable,
decay,francis,qflv,gyang,tany,qmeng,qqin,braaten}.

Recently, the LHCb Collaboration discovered the doubly charmed state
$T^+_{cc}$ with $IJ^P=01^+$ by analyzing the $D^0D^0\pi^+$ invariant
mass spectrum~\cite{tcc-exp1}, which has a minimal quark
configuration of $cc\bar{u}\bar{d}$. Its binding energy relative to
the $DD^*$ threshold and width are
\begin{eqnarray}
E_b=-273\pm61\pm5^{+11}_{-14}~\mbox{keV},\nonumber\\
\noalign{\smallskip}
\Gamma=410\pm165\pm43^{+18}_{-38}~\mbox{keV}.\nonumber
\end{eqnarray}
The LHCb Collaboration also released a more profound decay analysis,
in which the unitarized Breit-Wigner profile was
used~\cite{tcc-exp2}. Its binding energy and decay width were
updated as
\begin{eqnarray}
E_b=-361\pm40~\mbox{keV},~ \Gamma=47.8\pm1.9~\mbox{keV}.\nonumber
\end{eqnarray}
The binding energy and decay width of the $T^+_{cc}$ signal match
very well with the prediction of the $DD^\ast$ molecular state
\cite{tcc-lee,tcc-lmeng}. The discovery of the $X(3872)$
~\cite{x3872} pioneered the observation of a family of hidden-charm
and hidden-bottom tetraquark and pentaquark states in the past
decades. Similarly, the discovery of the $T^+_{cc}$ shall open a new
gate for a family of the $T^+_{cc}$-like doubly heavy tetraquark,
pentaquark and hexaquark states.

The discovery of the state $T^+_{cc}$ has inspired a large amount of
investigations on its properties and structure within the different
theoretical frameworks~\cite{tcc-oset,tcc-yan,tcc-fleming,tcc-nli,tcc-lmeng,tcc-twwu,
tcc-xzling,tcc-rchen,tcc-lydai,tcc-xzweng,tcc-qxin,tcc-rchen2,tcc-yhuang,
tcc-hren,tcc-yjin,tcc-yhu,tcc-albaladejo,tcc-oka}.
Many $T^+_{cc}$-like doubly heavy tetraquark candidates were
proposed, such as the states $T'_{cc}$~\cite{tcc-rchen},
$T^-_{bb}$~\cite{tcc-hren}, $T^+_{ccs}$~\cite{tccs,tccsbseq},
$T^0_{bc}$~\cite{tbclattice}, $T^0_{bcs}$~\cite{tbclattice},
$S$-wave $D_1D_1$, $D_1D_2^*$ and $D^*_2D_2^*$
states~\cite{tcc-flwang}, doubly charmed $P_{cc}$ and
triply charmed $H_{ccc}$ states~\cite{tcc-xkdong,pcc,hccc}.

In this work, we will analyze the underlying dynamics in the
formation of the loosely bound $T^+_{cc}$ state very carefully. We
will exhaust the most promising partners of the $T^+_{cc}$ and
compare the different binding mechanisms in the $T^+_{cc}$ and
$T^-_{bb}$ systems.

This paper is organized as follows. After the introduction, the
details of the quark model are given in Sec. II. The construction of
the wavefunctions of the doubly heavy tetraquark states with the
molecule configuration is shown in Sec. III. The numerical results
and discussions of the stable doubly heavy tetraquark states are
presented in the following sections. The last section is a brief
summary.

\section{quark model}

The underlying theory of strong interaction is quantum
Chromodynamics (QCD). At the hadronic scale, QCD is highly
non-perturbative due to the complicated infrared behavior of the
non-Abelian $SU(3)$ gauge group. At present it is still impossible
to derive the hadron spectrum analytically from the QCD Lagrangian.
The QCD-inspired constituent quark model remains a powerful tool in
obtaining physical insight for the complicated strong interaction
systems although the connection between the light current quarks in
QCD and the light constituent quarks in the quark model is not
established clearly.

The constituent quark model was formulated under the assumption that
the hadrons are color singlet nonrelativistic bound states of
constituent quarks with phenomenological effective masses and
various effective interactions. The model Hamiltonian used here can
be written as
\begin{eqnarray}
H_n & = & \sum_{i=1}^n \left(m_i+\frac{\mathbf{p}_i^2}{2m_i}
\right)-T_{c}+\sum_{i>j}^n V_{ij}, \nonumber\\
V_{ij}&=&V_{ij}^{oge}+V_{ij}^{obe}+V_{ij}^{\sigma}+V^{con}_{ij},\nonumber
\end{eqnarray}
where $m_i$ and $\mathbf{p}_i$ are the mass and momentum of the $i$-th
quark or antiquark, respectively. $T_c$ is the center-of-mass
kinetic energy of the states and should be deducted. $V_{ij}^{oge}$,
$V_{ij}^{obe}$, $V_{ij}^{\sigma}$, and $V^{con}_{ij}$ are the
one-gluon-exchange interaction, one-boson-exchange interaction
($\pi$, $K$ and $\eta$), $\sigma$-meson exchange interaction and
color confinement potential between the particles $i$ and $j$,
respectively.

The origin of the constituent quark mass can be traced back to the
spontaneous breaking of $SU(3)_L\otimes SU(3)_R$ chiral symmetry and
consequently constituent quarks should interact through the exchange
of Goldstone bosons \cite{chnqm}. Chiral symmetry breaking suggests
dividing quarks into two different sectors: light quarks ($u$, $d$
and $s$) where the chiral symmetry is spontaneously broken and heavy
quarks ($c$ and $b$) where the symmetry is explicitly broken. The
meson exchange interactions only occur in the light quark sector.
The central parts of the interactions originating from chiral
symmetry breaking can be resumed as follows~\cite{chiralmeson},
\begin{eqnarray}
V_{ij}^{obe} & = & V^{\pi}_{ij} \sum_{k=1}^3 \mathbf{F}_i^k
\mathbf{F}_j^k+V^{K}_{ij} \sum_{k=4}^7\mathbf{F}_i^k\mathbf{F}_j^k\nonumber\\
&+&V^{\eta}_{ij} (\mathbf{F}^8_i \mathbf{F}^8_j\cos\theta_P-\sin\theta_P),\nonumber\\
\noalign{\smallskip} V^{\chi}_{ij} & =
&\frac{g^2_{ch}}{4\pi}\frac{m^3_{\chi}}{12m_im_j}\frac{\Lambda^{2}_{\chi}}{\Lambda^{2}_{\chi}
-m_{\chi}^2}\boldsymbol{\sigma}_{i}\cdot
\boldsymbol{\sigma}_{j}\nonumber
\end{eqnarray}
\begin{eqnarray}
&\times&\left( Y(m_\chi r_{ij})-\frac{\Lambda^{3}_{\chi}}{m_{\chi}^3}Y(\Lambda_{\chi} r_{ij})\right),
~\chi=\pi,~K~\mbox{and}~\eta \nonumber \\
\noalign{\smallskip} V^{\sigma}_{ij} & = &-\frac{g^2_{ch}}{4\pi}
\frac{\Lambda^{2}_{\sigma}m_{\sigma}}{\Lambda^{2}_{\sigma}-m_{\sigma}^2}
\left( Y(m_\sigma
r_{ij})-\frac{\Lambda_{\sigma}}{m_{\sigma}}Y(\Lambda_{\sigma}r_{ij})
\right). \nonumber
\end{eqnarray}
The noncentral parts, tensor force and spin-orbit coupling, are not
given because we are only interested in the $S$-wave states here.
The function $Y(x)=\frac{e^{-x}}{x}$, $\mathbf{F}_{i}$ and
$\boldsymbol{\sigma}_{i}$ are the flavor $SU(3)$ Gell-Mann matrices and
spin $SU(2)$ Pauli matrices, respectively. $r_{ij}$ is the distance
between the particles $i$ and $j$. The mass parameters $m_{\pi}$,
$m_K$ and $m_{\eta}$ take their experimental values. The cutoff
parameters $\Lambda$s and the mixing angle $\theta_{P}$ take the
values from Ref.~\cite{chiralmeson}. The mass parameter $m_{\sigma}$
can be determined through the PCAC relation $m^2_{\sigma}\approx
m^2_{\pi}+4m^2_{u,d}$~\cite{masssigma}. The chiral coupling constant
$g_{ch}$ can be obtained from the $\pi NN$ coupling constant through
\begin{equation}
\frac{g_{ch}^2}{4\pi}=\left(\frac{3}{5}\right)^2\frac{g_{\pi
NN}^2}{4\pi}\frac{m_{u,d}^2}{m_N^2}.\nonumber
\end{equation}

Besides the chiral symmetry breaking, there also exists the
one-gluon-exchange (OGE) potential. From the non-relativistic
reduction of the OGE diagram in QCD for the point-like quarks, one
gets
\begin{eqnarray}
V_{ij}^{oge} & = &
{\frac{\alpha_{s}}{4}}\boldsymbol{\lambda}^c_{i}\cdot\boldsymbol{\lambda}_{j}^c\left({\frac{1}{r_{ij}}}-
{\frac{2\pi\delta(\mathbf{r}_{ij})\boldsymbol{\sigma}_{i}\cdot
\boldsymbol{\sigma}_{j}}{3m_im_j}}\right),\nonumber
\end{eqnarray}
$\boldsymbol{\lambda}_{i}$ is the color $SU(3)$ Gell-Mann matrices.
The Dirac $\delta(\mathbf{r}_{ij})$ function, where
$\mathbf{r}_{ij}=\mathbf{r}_{i}-\mathbf{r}_{j}$, arises from the
interaction between point-like quarks and collapses when not treated
perturbatively~\cite{collapse}. Therefore, the
$\delta(\mathbf{r}_{ij})$ function is regularized in the
form~\cite{chiralmeson}
\begin{equation}
\delta(\mathbf{r}_{ij})\rightarrow\frac{1}{4\pi
r_{ij}r_0^2(\mu_{ij})}e^{-r_{ij}/r_0(\mu_{ij})},\nonumber
\end{equation}
where $r_0(\mu_{ij})=\hat{r}_0/\mu_{ij}$, $\hat{r}_0$ is an
adjustable model parameter and $\mu_{ij}$ is the reduced mass of two
interacting particles $i$ and $j$. This regularization is justified
based on the finite size of the constituent quarks and should be
flavor dependent~\cite{flavor-dependent}.

The quark-gluon coupling constant $\alpha_s$ in the perturbative QCD
reads~\cite{alphas}
\begin{equation}
\alpha_s(\mu^2)=\frac{1}{\beta_0\ln\frac{\mu^2}{\Lambda^2}},\nonumber
\end{equation}
In the present work, we use an effective scale-dependent form given
by
\begin{equation}
\alpha_s(\mu^2_{ij})=\frac{\alpha_0}{\ln\frac{\mu_{ij}^2}{\Lambda_0^2}},\nonumber
\end{equation}
$\Lambda_0$ and $\alpha_0$ are adjustable model parameters
determined by fitting the ground state meson spectrum.

Finally, any model imitating QCD should incorporate the
nonperturbative color confinement effect. We adopt the
phenomenological color screening confinement potential,
\begin{eqnarray}
V^{con}_{ij}&=&-a_c\boldsymbol{\lambda}^c_i\cdot\boldsymbol{\lambda}^c_jf(r_{ij})\nonumber\\
\noalign{\smallskip} f(r_{ij})&=&\left \{
\begin{array}{cccccc}
r^2_{ij}~~~~~~~~~\mbox{if $i$, $j$ occur in the same meson}, \nonumber\\
\noalign{\smallskip} \frac{1-e^{-\mu_c r^2_{ij}}}{\mu_c} ~\mbox{if
$i$, $j$ occur in different mesons.}\nonumber
\end{array}
\right.
\end{eqnarray}
It is different from the form of confinement potential used in
recent investigations on the doubly heavy tetraquark states~\cite{gyang,tany}.
The adjustable parameter $a_c$ is determined by fitting the ground
state meson spectrum. The color screening parameter $\mu_c=1.0$
fm$^{-2}$ is taken from Ref. ~\cite{qdcsm}. The color screening
confinement potential can automatically match the quadratic one in
the short distance region ($\mu_cr^2\ll 1$). When two mesons are
separated to large distances, the confinement potential can
guarantee that the energy of the tetraquark system evolves into the
sum of the two-meson internal energy calculated by the model
Hamiltonian. In the intermediate range region, the hybrid
confinement can give a different picture from that given by a single
form confinement. Such type of the color screening confinement
potential comes from the quark delocalization and color screening
model~\cite{qdcsm}, which can describe the nuclear intermediate
range attraction and reproduce the $N$-$N$ scattering data and the
properties of the deuteron. Meanwhile, the model can avoid the
spurious van de Walls color force between two color singlets arising
from the direct extension of the single-hadron Hamiltonian to the
multiquark states~\cite{qdcsm}. The model has been widely applied to
investigate the properties of the baryon-baryon and baryon-meson
interactions~\cite{qdcsm1}.

\section{wave functions of the doubly heavy tetraquark states}

Our previous systematical investigation on the doubly heavy tetraquark states
suggested that the states $[Q_1Q_3][\bar{q}_2\bar{q}_4]$ can establish deep compact
bound states in the quark models~\cite{ours}, which is obviously contradictive
with the properties of the state $T^+_{cc}$ reported by the LHCb
Collaboration~\cite{tcc-exp1,tcc-exp2}.

The research on the states $[Q_1\bar{q}_2][Q_3\bar{q}_4]$ indicated that
their main color component is the molecule state $[Q_1\bar{q}_2]_{\mathbf{1}_c}
[Q_3\bar{q}_4]_{\mathbf{1}_c}$~\cite{gyang}. Furthermore, recent study on the
states $[c\bar{u}][c\bar{d}]$ and $[b\bar{u}][b\bar{d}]$ shown that the hidden
color components $\left[[c\bar{u}]_{\mathbf{8}_c} [c\bar{d}]_{\mathbf{8}_c}\right]_{\mathbf{1}_c}$
and $\left[[b\bar{u}]_{\mathbf{8}_c} [b\bar{d}]_{\mathbf{8}_c}\right]_{\mathbf{1}_c}$ can
be negligible if all possible color singlet components were taken into accounted~\cite{xychen}.
The influence of the hidden color components on the numerical results is very insignificant
even if the hidden color configurations are included. In order to reduce the heavy computational
workload, we therefore omit the hidden color configuration $\left[[Q_1\bar{q}_2]_{\mathbf{8}_c}
[Q_3\bar{q}_4]_{\mathbf{8}_c}\right]_{\mathbf{1}_c}$
in the present work.

Within the framework of the molecule configuration
$[c_1\bar{u}_2]_{\mathbf{1}_c}[c_3\bar{d}_4]_{\mathbf{1}_c}$,
the trial wave function of the $T^+_{cc}$ state with $IJ^P=01^+$
can be constructed as a sum of the following direct products of
color $\psi_c$, isospin $\eta_i$, spin $\chi_s$ and spatial $\phi$
terms
\begin{eqnarray}
\Phi^{T^+_{cc}}_{IJ}=\sum_{\alpha}\mathcal{A}\left
\{\left[\left[\phi_{l_am_a}^G(\mathbf{r})\chi_{s_a}\right]^{[c_1\bar{u}_2]}_{J_aM_{J_a}}
\left[\phi_{l_bm_b}^G(\mathbf{R})\chi_{s_b}\right]^{[c_3\bar{d}_4]}_{J_bM_{J_b}}\right.\right.\nonumber\\
\times\left.\left.\phi^G_{l_cm_c}(\boldsymbol{\rho})\right]^{T^+_{cc}}_{JM_J}\left[\eta_{i_a}^{[c_1\bar{u}_2]}
\eta_{i_b}^{[c_3\bar{d}_4]}\right]_{I}^{T^+_{cc}}\left[\psi_{c_a}^{[c_1\bar{u}_2]}\psi_{c_b}^
{[c_3\bar{d}_4]}\right]_{C}^{T^+_{cc}}\right\}. \nonumber
\end{eqnarray}
Here we assume the magnetic components $M_I=I$ and $M_J=J$. The
subscripts $a$ and $b$ represent the subclusters $[c_1\bar{u}_2]$
and $[c_3\bar{d}_4]$, respectively. $\mathcal{A}$ is the
antisymmetrization operator and equal to
$1-P_{13}-P_{24}+P_{13}P_{24}$ because of the Fermi-Dirac statistic
of the identical particles, where $P_{ij}$ is the permutation
operator on the particles $i$ and $j$. The summing index $\alpha$
stands for all possible flavor-spin-color-spatial intermediate
quantum numbers.

The relative spatial coordinates $\mathbf{r}$, $\mathbf{R}$ and
$\boldsymbol{\rho}$ are defined as
\begin{eqnarray}
\mathbf{r}&=&\mathbf{r}_1-\mathbf{r}_2,~\mathbf{R}=\mathbf{r}_3-\mathbf{r}_4,\nonumber\\
\noalign{\smallskip}
\boldsymbol{\rho}&=&\frac{m_1\mathbf{r}_1+m_2\mathbf{r}_2}{m_1+m_2}-\frac{m_3\mathbf{r}_3
+m_4\mathbf{r}_4}{m_3+m_4}.\nonumber
\end{eqnarray}
The corresponding angular excitations of three relative motions
are, respectively, $l_a$, $l_b$ and $l_c$. The parity of the state
$T^+_{cc}$ can therefore be expressed in terms of the relative
orbital angular momenta associated with the Jacobi coordinates as
$P=(-1)^{l_a+l_b+l_c}$. It is worth mentioning that this set of
coordinate is only a possible choice of many coordinates and however
most propitious to describe the correlation of two mesons. In order
to obtain a reliable solution of few-body problem, a high precision
numerical method is indispensable. The Gaussian Expansion Method
(GEM)~\cite{GEM} has been proven to be very powerful to solve
few-body problem. Brink et al first applied the GEM for studying
heavy tetraquarks states~\cite{constituentmodel1}. We also use
the GEM to study doubly heavy tetraquark systems in the present
work. According to the GEM, the relative motion wave function can
be written as
\begin{eqnarray}
\phi^G_{lm}(\mathbf{x})=\sum_{n=1}^{n_{max}}c_{n}N_{nl}x^{l}e^{-\nu_{n}x^2}
Y_{lm}(\hat{\mathbf{x}})\nonumber
\end{eqnarray}
Gaussian size parameters are taken as geometric progression
\begin{eqnarray}
\nu_{n}=\frac{1}{r^2_n},& r_n=r_1a^{n-1},&
a=\left(\frac{r_{n_{max}}}{r_1}\right)^{\frac{1}{n_{max}-1}}.
\end{eqnarray}
$r_1$ and $r_{max}$ are the minimum and maximum of the size,
respectively. $n_{max}$ is the number of the Gaussian wave function.
More details about the GEM can be found in Ref.~\cite{GEM}. In the
present work, we expand the wavefunction
$\phi_{l_am_a}^G(\mathbf{r})$ ($\phi_{l_bm_b}^G(\mathbf{R})$) with
$n_{max}$ ($n'_{max}$) Gaussian functions with the different width
ranging from 0.1 fm to 2.0 fm because the size of the mesons is less
than 1 fm. We expand the wavefunction
$\phi_{l_cm_c}^G(\boldsymbol{\rho})$ with $n''_{max}$ Gaussian functions
with the different width ranging from 0.1 fm to 5.0 fm because the
size of meson-meson molecule is about several fms. In this way, the
total number of the trial wave function $N_{base}$ is equal to
$n_{csf}\times n_{max}\times n'_{max}\times n''_{max}$, where
$n_{csf}$, the number of the color-spin-flavor wave function, will
be given later. $N_{base}$ should be increased gradually by
increasing $n_{max}$, $n'_{max}$ and $n''_{max}$ until the
convergent numerical results are obtained. In addition, a large
width Gaussian function $\phi_{l_cm_c}^G(\boldsymbol{\rho})$
($\rho=|\boldsymbol{\rho}|\rightarrow\infty$) should be introduced to
guarantee a fast convergence of numerical results when a bound state
does not exist.

Taking all degrees of freedom of identical particles, the Pauli
principle must be satisfied by imposing a restriction on the quantum
numbers of the mesons $c_1\bar{u}_2$ and $c_3\bar{d}_4$. The quantum
numbers must satisfy the relation $s_a+s_b-S+i_a+i_b-I+l_c=even$
when $s_a=s_b$ because the present boson system should satisfy the
Boson-Einstein statistic. According to the restriction, the $S$-wave
$(l_c=0)$ states with $00^+$ and $02^+$ do not exist. The wave
function of the $S$-wave state $[c_1\bar{u}_2][\bar{c}_3\bar{d}_4]$
with $01^+$ has two possible channels,
\begin{eqnarray}
\begin{array}{cccccc}
[DD^*]_{-}&=&\frac{1}{\sqrt{2}}(D^{0*}D^+-D^+D^*),~~~\nonumber \\
\noalign{\smallskip}\noalign{\smallskip}
[D^*D^*]_{-}&=&\frac{1}{\sqrt{2}}(D^{*0}D^{*+}-D^{*+}D^{*0}),\nonumber
\end{array}
\end{eqnarray}
In the channel $[D^*D^*]_{-}$, the spins of the mesons $D^{*0}$ and
$D^{*+}$ couple into the total angular momentum $J$. Similarly, the
wave function of the state $[c_1\bar{u}_2][\bar{c}_3\bar{d}_4]$ with
$10^+$ also has two possible channels,
\begin{eqnarray}
\begin{array}{cccccc}
[DD]_{+}&=&\frac{1}{\sqrt{2}}(D^0D^++D^+D^0),~~~~\nonumber\\
\noalign{\smallskip}\noalign{\smallskip}
[D^*D^*]_{+}&=&\frac{1}{\sqrt{2}}(D^{*0}D^{*+}+D^{*+}D^{*0}).\nonumber
\end{array}
\end{eqnarray}
The wave functions of the states
$[c_1\bar{u}_2][\bar{c}_3\bar{d}_4]$ with $11^+$ and $12^+$ can be
written as
\begin{eqnarray}
\begin{array}{cccccc}
[DD^*]_{+}&=&\frac{1}{\sqrt{2}}(D^0D^{*+}+D^{*+}D^0),~~ \nonumber\\
\noalign{\smallskip}\noalign{\smallskip}
[D^*D^*]_{+}&=&\frac{1}{\sqrt{2}}(D^{*0}D^{*+}+D^{*+}D^{*0}).\nonumber
\end{array}
\end{eqnarray}

Analogically, there may exist the partners of the $T^+_{cc}$ state
with the configurations of $[b_1\bar{u}_2][b_3\bar{d}_4]$ and
$[b_1\bar{u}_2][c_3\bar{d}_4]$, denoted as $T^-_{bb}$ and
$T^0_{bc}$. We can obtain the wave functions of the $S$-wave state
$T^-_{bb}$ by solely making a replacement of $c$ with $b$ in those
of the state $T^+_{cc}$. In the case of the state $T^0_{bc}$, the
wave functions of the states with $00^+$ and $10^+$ read
\begin{eqnarray}
\begin{array}{cccccc}
[\bar{B}D]_{\pm}&=&\frac{1}{\sqrt{2}}(B^-D^{+}\pm \bar{B}^{0}D^0),~~~\nonumber\\
\noalign{\smallskip}\noalign{\smallskip}
[\bar{B}^*D^*]_{\pm}&=&\frac{1}{\sqrt{2}}(B^{-*}D^{+*}\pm \bar{B}^{0*}D^{0*}).\nonumber
\end{array}
\end{eqnarray}
The signs $+$ and $-$ stand for the cases of the isospin symmetry
($I=1$) and antisymmetry ($I=0$), respectively. The same notations
hold for the other states with $I=1$ and $I=0$. The wave functions
of the states $[b_1\bar{u}_2][c_3\bar{d}_4]$ with $01^+$ and $11^+$
have the following three possible channels,
\begin{eqnarray}
\begin{array}{cccccc}
[\bar{B}D^*]_{\pm}&=&\frac{1}{\sqrt{2}}(B^-D^{+*}\pm \bar{B}^{0}D^{0*}),~\nonumber\\
\noalign{\smallskip}\noalign{\smallskip}
[\bar{B}^*D]_{\pm}&=&\frac{1}{\sqrt{2}}(B^{-*}D^{+}\pm \bar{B}^{0*}D^0),~~\nonumber\\
\noalign{\smallskip}\noalign{\smallskip}
[\bar{B}^*D^*]_{\pm}&=&\frac{1}{\sqrt{2}}(B^{-*}D^{+*}\pm \bar{B}^{0*}D^{0*}).\nonumber
\end{array}
\end{eqnarray}
The wave function of the states with $02^+$ and $12^+$ has the
following one possible channel,
\begin{eqnarray}
\begin{array}{cccccc}
[\bar{B}^*D^*]_{\pm}&=&\frac{1}{\sqrt{2}}(B^{-*}D^{+*}\pm \bar{B}^{0*}D^{0*}).\nonumber
\end{array}
\end{eqnarray}

We denote the strange partners of the $T^+_{cc}$ state with the
configurations $[b_1\bar{u}_2][b_3\bar{s}_4]$,
$[c_1\bar{u}_2][c_3\bar{s}_4]$ and $[b_1\bar{u}_2][c_3\bar{s}_4]$ as
the $T^-_{bbs}$, $T^+_{ccs}$ and $T^0_{bcs}$, respectively. We can
obtain their wave functions by replacing $\bar{d}$ with $\bar{s}$ in
those of the states $T^-_{bb}$, $T^+_{cc}$ and $T^0_{bc}$ because
$I$-spin (isospin) and $V$-spin are equivalent. The wave functions
of the states $[b_1\bar{s}_2][b_3\bar{s}_4]$,
$[c_1\bar{s}_2][c_3\bar{s}_4]$ and $[b_1\bar{s}_2][c_3\bar{s}_4]$,
denoted as the $T^{0}_{bbss}$, $T^{++}_{ccss}$ and $T^{+}_{bcss}$,
are similar to those of the states $T^-_{bb}$, $T^+_{cc}$ and $T^0_{bc}$
with $I=1$ because they have the same flavor symmetry.

\section{Structure of the $T^+_{cc}$ state with $01^+$}

We reproduce the mass spectrum of the ordinary mesons to determine
model parameters as in Ref.~\cite{ours}. We collect the results of
the heavy-light mesons in Table~\ref{mesons}.
\begin{table}
\caption{The $Q\bar{q}$ meson spectrum in the model where the mass
unit is in MeV and $\langle r^2\rangle^{\frac{1}{2}}$ unit in fm.}
\label{mesons}
\begin{tabular}{cccccccccccccccccc}
\toprule[0.8pt] \noalign{\smallskip}
~~State~~&$D$&~~~$D^*$~~~&$D_s$&~~~$D_s^*$~~~&$\bar{B}$&~~~$\bar{B}^*$&~~~$\bar{B}_s$&~~~$\bar{B}_s^*$~~~\\
\toprule[0.8pt] \noalign{\smallskip}
Cal.&1867&2002&1972&2140&5259&5301&5377&5430\\
\noalign{\smallskip}
PDG&1869&2007&1968&2112&5280&5325&5366&5416\\
$\langle r^2\rangle^{\frac{1}{2}}$&0.68&0.82&0.52&0.69&0.73&0.77&0.57&0.62\\
\toprule[0.8pt] \noalign{\smallskip}
\end{tabular}
\end{table}

In the following, we move on to the investigation of the $T^+_{cc}$
with $01^+$ and its partners. In order to obtain the lowest states
with positive parity, we assume that the three relative motions are
in $S$-wave, namely $l_a=l_b=l_c=0$. We can obtain the eigenvalue
and eigenvector by solving the four-body Schr\"{o}dinger equation
\begin{eqnarray}
(H_4-E_4)\Phi^{T^+_{cc}}_{IJ}=0\nonumber
\end{eqnarray}
with the Rayleigh-Ritz variational principle. We define the binding
energy $E_b$ of the doubly heavy tetraquark states as
\begin{eqnarray}
E_b=E_4-\lim_{\rho \rightarrow \infty}E_4(\rho)\nonumber
\end{eqnarray}
to identify whether or not the tetraquark states are stable against
the strong interactions, where $\lim_{\rho \rightarrow
\infty}E_4(\rho)$ is the lowest theoretical threshold of the two
mesons which can couple into the same quantum numbers with those of
the tetraquark states. Such a subtraction procedure can greatly
reduce the influence of the inaccurate model parameters and meson
spectra on the binding energies. If $E_b\geq0$, the tetraquark
systems can fall apart into two mesons via the strong interactions.
If $E_b<0$, the strong decay into two mesons is forbidden and
therefore the decay can only occur via either the weak or
electromagnetic interaction.
\begin{table*}[ht]
\caption{The stability of numerical results, $E_b$ unit in MeV, $r$,
$r'$, $r''$ and rms unit in fm.} \label{tcc}
\begin{tabular}{cccccccccccccccccc}
\toprule[0.8pt] \noalign{\smallskip}
&$\phi_{00}^G(\mathbf{r})$&&&$\phi_{00}^G(\mathbf{R})$&&&$\phi_{00}^G(\boldsymbol{\rho})$&&& $[DD^*]_{-}$ &$[D^*D^*]_{-}$&~~~Mixing~~~&&Rms&&\\
\noalign{\smallskip}
$r_1$&$r_{max}$&$n_{max}$&$r'_1$&$r'_{max}$&$n'_{max}$&$r''_1$&$r''_{max}$&$n''_{max}$&~~$N_{base}$~~&~~~$E_b$,~Ratio~~~&~~~$E_b$,~Ratio~~~&$E_b$&
$\langle\mathbf{r}^2\rangle^{\frac{1}{2}}$&~$\langle\mathbf{R}^2\rangle^{\frac{1}{2}}$~&$\langle\boldsymbol{\rho}^2\rangle^{\frac{1}{2}}$&\\
\noalign{\smallskip} \toprule[0.8pt] \noalign{\smallskip}
0.1&2.0&7   ~ &~0.1~&2.0&7  ~ &~0.1~&5.0&20  &1960 &0.000,~99.6\% &0.000,~0.4\%&\textcolor{red}{$-0.330$}&0.75&0.75&4.41   \\
\noalign{\smallskip}
0.1&2.0&8   ~ &~0.1~&2.0&8  ~ &~0.1~&5.0&21  &2688 &0.000,~99.6\% &0.000,~0.4\%&\textcolor{red}{$-0.353$}&0.75&0.75&4.23   \\
\noalign{\smallskip}
0.1&2.0&9   ~ &~0.1~&2.0&9  ~ &~0.1~&5.0&22  &3654 &0.000,~99.6\% &0.000,~0.4\%&\textcolor{red}{$-0.344$}&0.75&0.75&4.30   \\
\noalign{\smallskip}
0.1&2.0&10  ~ &~0.1~&2.0&10 ~ &~0.1~&5.0&23  &4600 &0.000,~99.6\% &0.000,~0.4\%&\textcolor{red}{$-0.342$}&0.75&0.75&4.32   \\
\noalign{\smallskip}
0.1&2.0&11  ~ &~0.1~&2.0&11 ~ &~0.1~&5.0&24  &5808 &0.000,~99.6\% &0.000,~0.4\%&\textcolor{red}{$-0.342$}&0.75&0.75&4.32  \\
\noalign{\smallskip}
0.1&2.0&12  ~ &~0.1~&2.0&12 ~ &~0.1~&5.0&25  &7200 &0.000,~99.6\% &0.000,~0.4\%&\textcolor{red}{$-0.342$}&0.75&0.75&4.32   \\
\noalign{\smallskip} \toprule[0.8pt]
\end{tabular}
\end{table*}

\begin{table*}
\caption{The properties of the state $T^+_{cc}$ and its partners
predicted by the model, $E_b$ unit in MeV and rms unit in fm.
$\Delta E_k$ is the kinetic energy difference between the tetraquark
system and its corresponding threshold. Four cases: (a) with meson
exchange and color screening effect; (b) with meson exchange while
without color screening effect; (c) without meson exchange while
with color screening effect; (d) without meson exchange and color
screening effect. The ``$-$" denotes that the single channel is
unbound.} \label{partners}
\begin{tabular}{ccccccccccccccccccccccc}
\toprule[0.8pt] \noalign{\smallskip}
State&$IJ^P$&&Channel&&Mixing&&Rms&&&&&$E_b^i$&&&\\
Case&$\mu_c$&&$E_b$,~ratio&&$E_b$&$\langle\mathbf{r}^2\rangle^{\frac{1}{2}}$&$\langle\mathbf{R}^2\rangle^{\frac{1}{2}}$&
$\langle\boldsymbol{\rho}^2\rangle^{\frac{1}{2}}$&$V^{\sigma}$&$ V^{\pi}$&~$V^{K}$~&$V^{\eta}$&$V^{con}$&$V^{oge}$&$\Delta E_k$\\
\noalign{\smallskip} \toprule[0.8pt] \noalign{\smallskip}
$T^+_{cc}$&\textcolor{red}{$01^+$}&&$[\bar{D}\bar{D}^*]_{-}$&$[\bar{D}^*\bar{D}^*]_{-}$&\\
\noalign{\smallskip}
(a)&$1.0~$&&$\mathbf{-}$,~$99.6\%$&$\mathbf{-}$,~$0.4\%$&\textcolor{red}{$-0.34$}&0.75&0.75&4.32 &$-2.68$&$-2.19$&$0.00$&$0.17$&$-1.00$&$-4.29$&$9.65$  \\
\noalign{\smallskip}
(b)&$0.0~$&&$\mathbf{-}$,~$98.8\%$&$\mathbf{-}$,~$1.2\%$&\textcolor{red}{$-0.86$}&0.76&0.76&2.94 &$-3.72$&$-5.41$&$0.00$&$0.49$&$-1.87$&$-7.88$&$17.51$  \\
\noalign{\smallskip}
(c)&$1.0~$&&$\mathbf{-}$,~$100\%$&$\mathbf{-}$,~$0.0\%$&$~~\mathbf{-}$&  \\
\noalign{\smallskip}
(d)&$0.0~$&&$\mathbf{-}$,~$100\%$&$\mathbf{-}$,~$0.0\%$&$~~\mathbf{-}$&  \\
\noalign{\smallskip} \toprule[0.8pt] \noalign{\smallskip}
$T^-_{bb}$&\textcolor{red}{$01^+$}&&$[\bar{B}\bar{B}^*]_{-}$&$[\bar{B}^*\bar{B}^*]_{-}$\\
\noalign{\smallskip}
(a)&$1.0~$&&$\mathbf{-11.2}$,~$80.6\%$&$\mathbf{-9.8}$,~$19.4\%$&\textcolor{red}{$-28.6$}&0.79&0.79&0.59 &$-15.8$&$-57.9$&$0.0$&$6.6$&$-16.7$&$-108.3$&$163.5$  \\
\noalign{\smallskip}
(b)&$0.0~$&&$\mathbf{-10.0}$,~$62.4\%$&$\mathbf{-9.0}$,~$37.6\%$&\textcolor{red}{$-43.8$}&0.84&0.84&0.46 &$-17.0$&$-76.0$&$0.0$&$8.6$&$-21.8$&$-111.7$&$174.1$  \\
\noalign{\smallskip}
(c)&$1.0~$&&$\mathbf{-0.3}$,~$84.7\%$&$\mathbf{-0.2}$,~$15.3\%$&\textcolor{red}{$-10.0$}&0.72&0.72&1.04&&&&&$-2.5$&3.4&$-10.9$  \\
\noalign{\smallskip}
(d)&$0.0~$&&$\mathbf{-}$,~$94.8\%$&$\mathbf{-}$,~$5.2\%$&\textcolor{red}{$-3.9$}&0.73&0.73&1.65&&&&&$-1.4$&$-10.1$&$7.6$  \\
\noalign{\smallskip} \toprule[0.8pt] \noalign{\smallskip}
$T^0_{bc}$&\textcolor{red}{$00^+$}&&[$\bar{B}D]_{-}$&$[\bar{B}^*D^*]_{-}$\\
\noalign{\smallskip}
(a)&$1.0~$&&$\mathbf{-9.0}$,~$98.6\%$&$\mathbf{-}$,~$1.4\%$&\textcolor{red}{$-12.9$} &0.71&0.66&1.09&$-9.9$&$-1.4$&0.0&$0.2$&$-5.1$&$-28.6$&31.9 \\
\noalign{\smallskip}
(b)&$0.0~$&&$\mathbf{-5.7}$,~$98.6\%$&$\mathbf{-3.5}$,~$1.4\%$&\textcolor{red}{$-10.5$} &0.71&0.67&1.14&$-9.9$&$-5.0$&0.0&$0.6$&$-4.9$&$-34.4$&43.1 \\
\noalign{\smallskip}
(c)&$1.0~$&&$\mathbf{-4.8}$,~$99.6\%$&$\mathbf{-}$,~$0.4\%$&\textcolor{red}{$-6.5$}&0.71&0.66&1.37&&&&&$-2.3$&$-9.2$&5.0 \\
\noalign{\smallskip}
(d)&$0.0~$&&$\mathbf{-2.1}$,~$99.8\%$&$\mathbf{-}$,~$0.2\%$&\textcolor{red}{$-3.0$}&0.71&0.66&1.82&&&&&$-1.2$&$-13.1$&11.4 \\
\noalign{\smallskip}
$T^0_{bc}$&\textcolor{red}{$01^+$}&$[\bar{B}D^*]_{-}$&$[\bar{B}^*D]_{-}$&$[\bar{B}^*D^*]_{-}$\\
\noalign{\smallskip}
(a)&$1.0~$&$\mathbf{-6.5}$,~$0.5\%$&$\mathbf{-7.7}$,~$98.7\%$&$\mathbf{-3.0}$,~$0.8\%$&\textcolor{red}{$-10.5$}&0.75&0.66&1.20&$-8.4$&$0.6$&0.0&$0.1$&$-4.0$&$-15.7$&16.9    \\
\noalign{\smallskip}
(b)&$0.0~$&$\mathbf{-2.7}$,~$0.1\%$&$\mathbf{-4.4}$,~$97.2\%$&$\mathbf{-1.3}$,~$2.7\%$&\textcolor{red}{$-7.6$}&0.77&0.68&1.30&$-8.1$&$-3.6$&0.0&$0.4$&$-3.8$&$-21.7$&29.2    \\
\noalign{\smallskip}
(c)&$1.0~$&$\mathbf{-3.5}$,~$0.8\%$&$\mathbf{-3.9}$,~$98.7\%$&$\mathbf{-}$,~$0.5\%$&\textcolor{red}{$-6.3$}&0.75&0.66&1.40&&&&&$-2.1$&$-3.2$&$-1.1$    \\
\noalign{\smallskip}
(d)&$0.0~$&$\mathbf{-0.6}$,~$0.2\%$&$\mathbf{-1.4}$,~$99.6\%$&$\mathbf{-}$,~$0.2\%$&\textcolor{red}{$-2.2$}&0.76&0.67&2.07&&&&&$-0.9$&$-8.0$&$6.7$    \\
\noalign{\smallskip}
$T^0_{bc}$&\textcolor{red}{$02^+$}&&&$[\bar{B}^*D^*]_{-}$&\\
\noalign{\smallskip}
(a)&$1.0~$&&&$\mathbf{-12.0}$,~$100\%$&\textcolor{red}{$-12.0$}&0.74&0.79&1.28&$-6.4$&$7.3$&$0.0$&$-0.5$&$-2.8$&$0.4$&$-10.0$\\
\noalign{\smallskip}
(b)&$0.0~$&&&$\mathbf{-3.6}$,~$100\%$&\textcolor{red}{$-3.6$}&0.75&0.79&1.88&$-4.3$&$4.5$&$0.0$&$-0.3$&$-0.8$&$-3.7$&$0.9$\\
\noalign{\smallskip}
(c)&$1.0~$&&&$\mathbf{-12.4}$,~$100\%$&\textcolor{red}{$-12.4$}&0.74&0.79&1.25&&&&&$-3.7$&$-0.6$&$-8.0$\\
\noalign{\smallskip}
(d)&$0.0~$&&&$\mathbf{-3.6}$,~$100\%$&\textcolor{red}{$-3.6$}&0.75&0.79&1.83&&&&&$-1.3$&$-4.6$&$2.0$\\
\noalign{\smallskip} \toprule[0.8pt] \noalign{\smallskip}
$T^-_{bbs}$&\textcolor{red}{$\frac{1}{2}1^+$}&&$[\bar{B}\bar{B}_s^*]_-$&$[\bar{B}^*\bar{B}_s^*]_-$\\
\noalign{\smallskip}
(a)&$1.0~$&&$\mathbf{-4.8}$,~$91.1\%$&$\mathbf{-4.2}$,~$8.9\%$&\textcolor{red}{$-11.8$}&0.66&0.66&0.94&$-11.2$&0.0&0.3&0.1&$-3.5$&$-6.1$&$8.6$\\
\noalign{\smallskip}
(b)&$0.0~$&&$\mathbf{-3.8}$,~$96.0\%$&$\mathbf{-3.4}$,~$4.0\%$&\textcolor{red}{$-9.0$}&0.68&0.68&0.97&$-11.4$&0.0&$-2.3$&$-0.4$&$-4.0$&$-21.4$&$30.4$\\
\noalign{\smallskip}
(c)&$1.0~$&&$\mathbf{-}$,~$90.1\%$&$\mathbf{-}$,~$9.9\%$&\textcolor{red}{$-3.3$}&0.67&0.67&1.30&&&&&$-0.6$&$5.4$&$-8.1$\\
\noalign{\smallskip}
(d)&$0.0~$&&$\mathbf{-}$,~$97.6\%$&$\mathbf{-}$,~$2.4\%$&\textcolor{red}{$-0.5$}&0.68&0.68&2.74&&&&&$-0.2$&$-1.6$&$1.3$\\
\noalign{\smallskip} \toprule[0.8pt] \noalign{\smallskip}
$T^0_{bcs}$&\textcolor{red}{$\frac{1}{2}0^+$}&&$[\bar{B}_s D]_-$&$[{B}^*_s D^*]_-$\\
\noalign{\smallskip}
(a)&$1.0~$&&$\mathbf{-7.4}$,~$99.6\%$&$\mathbf{-}$,~$0.4\%$&\textcolor{red}{$-9.2$}&0.64&0.60&1.13&$-10.3$&0.0&0.8&0.1&$-3.0$&$-15.6$&18.8\\
\noalign{\smallskip}
(b)&$0.0~$&&$\mathbf{-5.1}$,~$99.6\%$&$\mathbf{-}$,~$0.4\%$&\textcolor{red}{$-6.7$}&0.64&0.60&1.25&$-9.5$&0.0&0.4&0.1&$-2.4$&$-20.2$&24.9\\
\noalign{\smallskip}
(c)&$1.0~$&&$\mathbf{-1.1}$,~$99.8\%$&$\mathbf{-}$,~$0.2\%$&\textcolor{red}{$-1.6$}&0.65&0.60&2.00&&&&&0.6&$-4.3$&2.1\\
\noalign{\smallskip}
(d)&$0.0~$&&$\mathbf{0.1}$,~$99.9\%$&$\mathbf{-}$,~$0.1\%$&\textcolor{red}{$-0.4$}&0.65&0.61&3.33&&&&&$-0.2$&$-4.2$&4.0\\
\noalign{\smallskip}
$T^0_{bcs}$&\textcolor{red}{$\frac{1}{2}1^+$}&$[\bar{B}_s D^*]_-$&$[\bar{B}^*_sD]_-$&$[\bar{B}_s^*D^*]_-$\\
\noalign{\smallskip}
(a)&$1.0~$&$\mathbf{-4.9}$,~$0.5\%$&$\mathbf{-6.1}$,~$99.1\%$&$\mathbf{-}$,~$0.4\%$&\textcolor{red}{$-8.0$}&0.69&0.60&1.21&$-8.9$&0.0&1.0&0.2&$-2.5$&$-6.7$&8.9\\
\noalign{\smallskip}
(b)&$0.0~$&$\mathbf{-2.3}$,~$0.2\%$&$\mathbf{-3.8}$,~$99.6\%$&$\mathbf{-}$,~$0.2\%$&\textcolor{red}{$-5.0$}&0.69&0.60&1.42&$-7.8$&0.0&0.6&0.1&$-1.8$&$-11.9$&15.8\\
\noalign{\smallskip}
(c)&$1.0~$&$\mathbf{-0.5}$,~$0.4\%$&$\mathbf{-0.7}$,~$99.4\%$&$\mathbf{-}$,~$0.2\%$&\textcolor{red}{$-1.8$}&0.70&0.60&2.04&&&&&$-0.5$&$-2.1$&$0.9$\\
\noalign{\smallskip}
(d)&$0.0~$&$\mathbf{-}$,~$0.1\%$&$\mathbf{-}$,~$99.8\%$&$\mathbf{-}$,~$0.1\%$&\textcolor{red}{$-0.2$}&0.70&0.60&5.53&&&&&$-0.1$&$-2.2$&$2.1$\\
\noalign{\smallskip}
$T^0_{bcs}$&\textcolor{red}{$\frac{1}{2}2^+$}&&&$[\bar{B}_s^*D^*]_-$&\\
\noalign{\smallskip}
(a)&$1.0~$&&&$\mathbf{-12.2}$,~$100\%$&\textcolor{red}{$-12.2$}&0.68&0.73&1.15&$-7.9$&0.0&1.7&0.3&$-3.2$&$-0.2$&$-2.8$\\
\noalign{\smallskip}
(b)&$0.0~$&&&$\mathbf{-5.4}$,~$100\%$&\textcolor{red}{$-5.4$}&0.68&0.73&1.48&$-6.2$&0.0&1.2&0.2&$-1.7$&$-4.6$&$5.7$\\
\noalign{\smallskip}
(c)&$1.0~$&&&$\mathbf{-6.6}$,~$100\%$&\textcolor{red}{$-6.6$}&0.69&0.75&1.36&&&&&$-1.6$&1.3&$-6.4$\\
\noalign{\smallskip}
(d)&$0.0~$&&&$\mathbf{-1.4}$,~$100\%$&\textcolor{red}{$-1.4$}&0.69&0.75&2.33&&&&&$-0.4$&$-1.9$&$1.0$\\
\noalign{\smallskip} \toprule[0.8pt]
\end{tabular}
\end{table*}

We can obtain the convergent numerical results of the state $T^+_{cc}$
with $01^+$ with more than 4600 bases in the
spin-color-flavor-orbital space, which are presented in
Table~\ref{tcc}. It can be seen from Table~\ref{tcc} that the
binding energy $E_b=-0.342$ MeV predicted by the model is highly
consistent with the data given by the LHCb Collaboration. Note that
no adjustable parameter is introduced to match the experimental data
in our calculation. Neither of the single $[DD^*]_-$ and
$[D^*D^*]_-$ channel alone can form a bound state in the model. The
stable $T^+_{cc}$ state arises from  the coupling of these two
channels. The dominant component of the state $T^+_{cc}$ is the $DD^*$
channel. Therefore, the coupled channel effect plays a critical role
in the formation of the state $T^+_{cc}$ under the assumption of the
molecule picture.

The spatial configuration of the state $T^+_{cc}$ can be ascertained
by the sizes of the two subclusters $D$ ($D^*$) and $D^*$ and their
relative distance, which can be approximately described by the rms
$\langle\mathbf{r}^2\rangle^{\frac{1}{2}}$,
$\langle\mathbf{R}^2\rangle^{\frac{1}{2}}$ and
$\langle\boldsymbol{\rho}^2\rangle^{\frac{1}{2}}$ determined by the
eigenvectors, respectively. The sizes
$\langle\mathbf{r}^2\rangle^{\frac{1}{2}}$ and
$\langle\mathbf{R}^2\rangle^{\frac{1}{2}}$ are approximately those
of the individual meson listed in Table~\ref{mesons}, which is far
less than the distance between two subclusters
$\langle\boldsymbol{\rho}^2\rangle^{\frac{1}{2}}= 4.32$ fm. In other
words, the two subclusters are far away from each other. Therefore,
the $T^+_{cc}$ state is a loosely bound deuteron-like state
consisted of $D$ and $D^*$, see Fig. 1. The extracted $T^+_{cc}$ size
4.32 fm confirms the prediction of a huge size 4.46 fm with a binding
energy 0.47 MeV in Ref. \cite{tcc-lee}, which also agrees with the spatial
configuration given by the LHCb Collaboration according to the
characteristic size calculated from the binding
energy~\cite{tcc-exp2}.

In order to illustrate the mechanism of the formation of the
$T^+_{cc}$ state, we calculate and decompose the contribution to the
binding energy $E_b$ from various parts of the Hamiltonian $E_b^i$
in the following four cases: (a) with meson exchange and color
screening effect, $\mu_c=1$; (b) with meson exchange while without
color screening effect, $\mu_c=0$; (c) without meson exchange while
with color screening effect, $\mu_c=1$; (d) without meson exchange
and color screening effect, $\mu_c=0$. In the present model, the
meson exchange and color screening effect only occur between two
subclusters in the $T^+_{cc}$ state and do not affect the
corresponding threshold. We present the numerical results in
Table~\ref{partners}, in which $\Delta E_k$ is the kinetic energy
difference between the tetraquark system and its corresponding
threshold. In addition, the binding energy of each single channel
and its ratio in the eigenvector are also given in
Table~\ref{partners}.

The hybrid color screening confinement potential generally gives
bigger binding energies than single type of one, see the cases (a)
and (b) or (c) and (d) in the states $T^0_{bc}$, $T^-_{bbs}$ and
$T^0_{bcs}$. However, the order is reversed in the cases (a) and (b)
of the states $T^+_{cc}$ and $T^-_{bb}$ because of their stronger
meson exchange interaction relative to the confinement potential.

\begin{figure} [h]
\resizebox{0.48\textwidth}{!}{\includegraphics{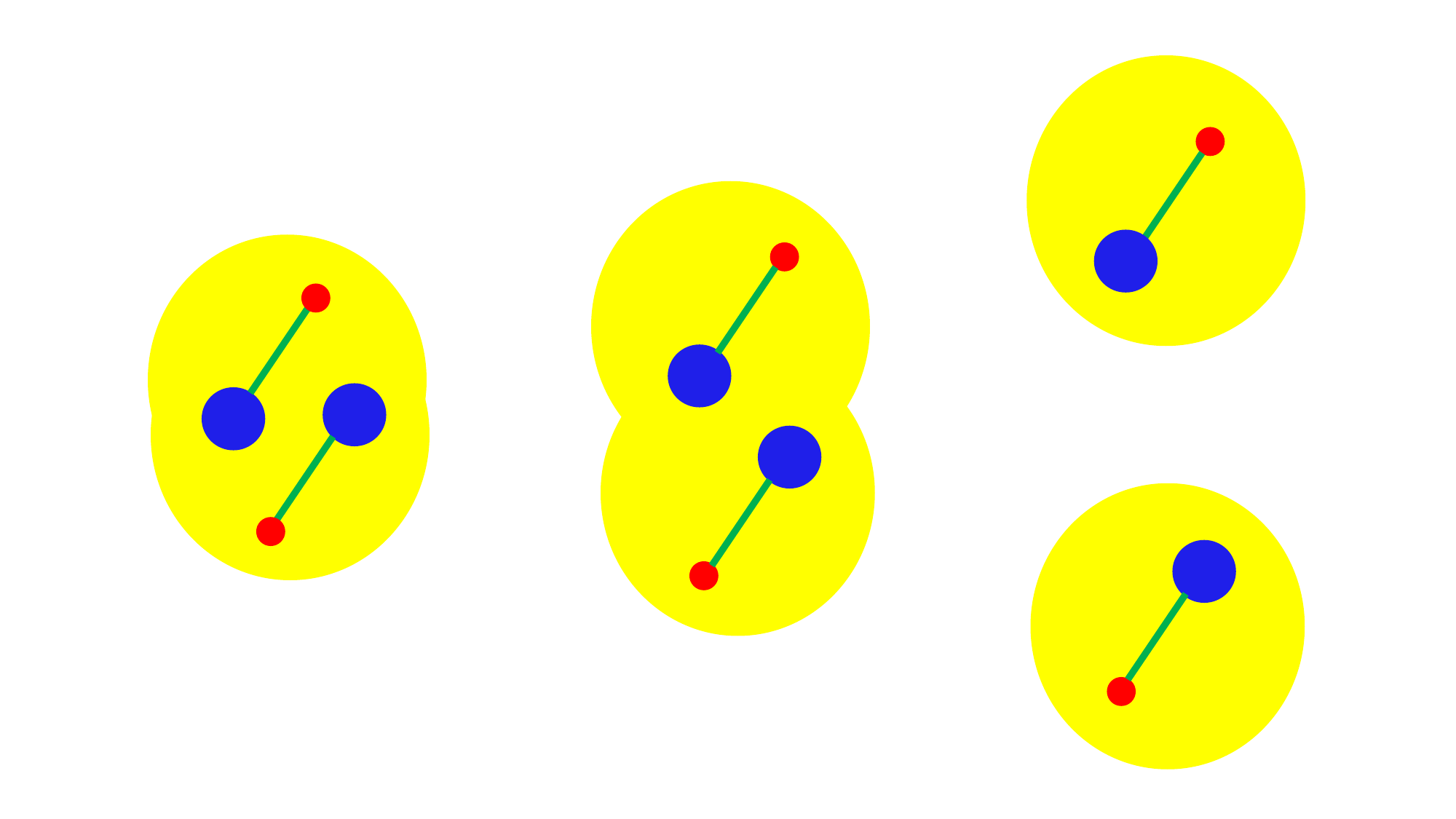}}
\caption{The spatial configuration, left: compact state;
middle: hydrogen molecule-like state; right: deuteron-like state,
 big blue ball and small red ball represent $Q$ and $\bar{q}$, respectively.
 A large yellow ball represents a subcluster.}
\label{2q-3q}
\end{figure}

From the cases (c) and (d) in Table~\ref{partners}, the bound state
$T^+_{cc}$ vanishes if we turn off the meson exchange interactions
in the model. In other words, the long-range $\pi$ and
intermediate-range $\sigma$ meson exchange force play a pivotal role
in the formation of the loosely bound $T^+_{cc}$ state.

We further compare the cases (a) and (b) to illustrate the very delicate
competition between the kinetic energy and the attraction from
various sources in the Hamiltonian. The main factor hindering the
formation of the $D$ ($D^*$) and $D^*$ subclusters into the bound
state is the relative kinetic energy between two subclusters in the
cases (a) and (b). The interactions $V^{\sigma}$, $ V^{\pi}$,
$V^{con}$ and $V^{oge}$ provide precious attractions. Without the
color screening effect, the confinement and one-gluon-exchange color
forces between the two subclusters become stronger, which pull them
closer to each other. Now all the meson exchange contributions
become larger in magnitude. Especially, the one-pion-exchange force
is extremely sensitive to the distance and its contribution to the
binding energy increases to $-5.41$ MeV. In contrast, the kinetic
energy difference $\Delta E_k$ increases to 17.51 MeV in the case (b).
The binding energy $E_b=-0.86$ MeV in the case (b) is also in good
agreement with the experimental data. Therefore, the $T^+_{cc}$
state always emerges as a loosely bound state so long as there
exists the meson exchange interactions.

\section{Isospin antisymmetric $T^-_{bb}$ states}

\subsection{Compact $T^-_{bb}$}

Both of the $[\bar{B}\bar{B}^*]_-$ and $[\bar{B}^*\bar{B}^*]_-$
channels with $01^+$ can form a bound state alone and their binding
energies are about 10 MeV in the cases (a) and (b). After coupling the
two channels, the state $T^-_{bb}$ becomes a rather deeply bound
state with $E_b=-28.6$ MeV and $-43.8$ MeV, respectively, which are
much less that of the bound state $T^-_{bb}$ with diquark-antidiquark
structure $[bb][\bar{u}\bar{d}]$ because of their different color
configurations~\cite{ours,qmeng}. Our present results are very close
to the latest lattice QCD predictions in the range of 20-40
MeV~\cite{tbblattice}. However, the earlier lattice QCD results
indicated that these states were over 100 MeV below the
$\bar{B}\bar{B}^*$ threshold~\cite{tbblattice2,tbblattice3}.
The strong attraction comes from the interactions $V^{\sigma}$,
$V^{\pi}$, $V^{con}$ and $V^{oge}$ in the model. Especially, the
contributions from the $V^{oge}$ and $V^{\pi}$ are quite large,
which is due to the rather compact size of the $T^-_{bb}$ system.

The main component of the $T^-_{bb}$ is the $[\bar{B}\bar{B}^*]_-$.
The two subclusters $\bar{B}$ ($\bar{B}^*$) and $\bar{B}^*$ become
obscure and overlap with each other severely because the sizes of
the subclusters $\langle\mathbf{r}^2\rangle^{\frac{1}{2}}$ and
$\langle\mathbf{R}^2\rangle^{\frac{1}{2}}$ are bigger than the
relative distance $\langle\boldsymbol{\rho}^2\rangle^{\frac{1}{2}}$
between the two subclusters, see Fig. 1. The large $b$ quark mass allows
the two subclusters to get as close as possible. The $T^-_{bb}$ state
with $01^+$ looks like a compact tetraquark state if there exists the
meson exchanges interaction. If so, the $T^-_{bb}$ state may not be a
pure meson-meson molecule state. Instead, it may be a mixture of the
meson-meson molecule and other hidden color states. The specific
ratio between two components needs further study. Such a qualitative
feature is supported by the lattice QCD
computations~\cite{tbblattice,tbblattice4}, in which the ratio of
the meson-meson molecule component is about 60\%.

\subsection{Deuteron-like $T^-_{bb}$}

If we remove the meson exchange interactions and color screening
effect from the model and focus on the case (d) in Table~\ref{partners},
the $T^-_{bb}$ state with $01^+$ becomes a shallow bound state with
$E_b=-3.9$ MeV and
$\langle\boldsymbol{\rho}^2\rangle^{\frac{1}{2}}=1.65$ fm, where the
attraction mainly comes from the residual one-gluon-exchange
potential $V^{oge}$. The two subclusters are separated too far away
to overlap each other because the sum of the sizes of the
subclusters $\langle\mathbf{r}^2\rangle^{\frac{1}{2}}$ and
$\langle\mathbf{R}^2\rangle^{\frac{1}{2}}$ is less than their
relative distance $\langle\boldsymbol{\rho}^2\rangle^{\frac{1}{2}}$.
Quarks are only allowed to move in the isolated subclusters.
Therefore, the $T^-_{bb}$ state with $01^+$ looks like a loosely
bound deuteron-like molecular state in the case (d), see Fig. 1.

\subsection{Hydrogen molecule-like $T^-_{bb}$}

If we consider the color screening effect in the case (c) in Table III,
the $T^-_{bb}$ state with $01^+$ forms a bound state with $E_b=-10$
MeV and $\langle\boldsymbol{\rho}^2\rangle^{\frac{1}{2}}=1.04$ fm, where
the attraction mainly arises from the decreasing of the kinetic
energy. This novel mechanism is completely different from those in
the other three cases, where the $V^{oge}$ and (or) the meson
exchange interactions provide a strong attraction while the kinetic
energy prevents the two subclusters to form a bound state.

The two subclusters overlap with each other extremely in the cases (a)
and (b) while they do not overlap at all in the case (d). Now in the case
(c), the two subclusters $\bar{B}$ ($\bar{B}^*$) and $\bar{B}^*$
moderately overlap with each other,  see Fig. 1. Such an appropriate
spatial overlapping greatly enlarges the phase space of the light
quarks $\bar{q}_2$ and $\bar{q}_4$ and allows them to roam into the
opposite subcluster freely, which helps to lower the kinetic energy
of the $T^-_{bb}$ system. This is the realization of the uncertainty
principle.

The delocalization of the light quarks in the state $T^-_{bb}$ is
extremely similar to the valence bond in the hydrogen molecule,
where the electron pair is shared by two protons. Therefore, the
$T^-_{bb}$ state with $01^+$ is very similar to the hydrogen
molecule state, which is formed by the delocalization of the light
quarks with the color screening effect in the case (c) in the present
model. The idea of the QCD valence bond was proposed and
investigated in Ref. \cite{qcdsum} in 2013 and discussed extensively
in the review \cite{tbb-zhu}. Recently, Maiani et al discussed the
hydrogen molecule-like $T^-_{bb}$ state when the $[bb]$ pair is in
color $\mathbf{6}$~\cite{h2-like}. Richard et al also studied the
hydrogen molecule-like doubly heavy tetraquark states~\cite{h2-like2}.

\subsection{Helium-like QCD atom in the limit of a large $m_Q$}

In order to reveal the dependence of the three configurations on the
heavy quark mass, we increase the bottom quark mass from $m_b$ to
$m_Q$ with the mass ratio $\frac{m_Q}{m_b}$ and calculate the
binding energy $E_b$ and the average distances. We present numerical
results in Table~\ref{variation}. One can see that the binding
energy $E_b$ in the three configurations is very sensitive to the
mass ratio. The deeply bound state appears in the limit of a large
$m_Q$. The large heavy quark mass permits them to get as close as
possible (see $\langle\mathbf{r}_{QQ}^2\rangle^{\frac{1}{2}}$ in
Table~\ref{variation}), therefore their attractive Coulomb
interaction becomes dominant. The binding energies of the B
(deuteron-like) and C (hydrogen-like) configurations are close to
each other. Their absolute values are smaller than that of the A
(compact) configuration because of the extra attraction from the
meson exchange interactions.

\begin{table}[h]
\caption{Variation of the configurations with the mass ratio $\frac{m_Q}{m_b}$.
$\langle\mathbf{r}_{QQ}^2\rangle^{\frac{1}{2}}$ and $\langle\mathbf{r}_{\bar{u}\bar{d}}^2\rangle^{\frac{1}{2}}$
are the average distance between $Q$ and $Q$ and $\bar{u}$ and $\bar{d}$, respectively.
Others have their original meanings.}\label{variation}
\begin{tabular}{cccccccccccccccccc}
\toprule[0.8pt] \noalign{\smallskip}
Conf.&~$\frac{m_Q}{m_b}$~&~~~$E_b$~~~&$\langle\mathbf{r}_{QQ}^2\rangle^{\frac{1}{2}}$
&$\langle\mathbf{r}_{\bar{u}\bar{d}}^2\rangle^{\frac{1}{2}}$&~$\langle\boldsymbol{\rho}^2\rangle^{\frac{1}{2}}$~&~
$\langle\mathbf{r}^2\rangle^{\frac{1}{2}}~$
&$\langle\mathbf{R}^2\rangle^{\frac{1}{2}}$\\
\noalign{\smallskip} \toprule[0.8pt] \noalign{\smallskip}
&$1$&$-28.6$&0.62&1.13&0.59&0.79&0.79\\
\noalign{\smallskip}
&$2$&$-93.5$&0.26&0.97&0.26&0.81&0.81\\
\noalign{\smallskip}
A&$3$&$-140.6$&0.19&0.95&0.19&0.81&0.81\\
\noalign{\smallskip}
&$4$&$-190.0$&0.16&0.94&0.16&0.81&0.81\\
\noalign{\smallskip}
&$5$&$-226.0$&0.14&0.93&0.14&0.81&0.81\\
\noalign{\smallskip}
\toprule[0.8pt]
\noalign{\smallskip}
&$1$&$-3.9$&1.73&1.95&1.65&0.73&0.73\\
\noalign{\smallskip}
&$2$&$-19.2$&0.65&1.19&0.62&0.80&0.80\\
\noalign{\smallskip}
B&$3$&$-58.1$&0.22&1.06&0.22&0.91&0.91\\
\noalign{\smallskip}
&$4$&$-97.1$&0.17&1.06&0.17&0.91&0.91\\
\noalign{\smallskip}
&$5$&$-131.4$&0.15&1.06&0.15&0.91&0.91\\
\toprule[0.8pt] \noalign{\smallskip}
\noalign{\smallskip}
&$1$&$-10.0$&1.10&1.48&1.04&0.72&0.72\\
\noalign{\smallskip}
&$2$&$-25.9$&0.79&1.29&0.77&0.78&0.78\\
\noalign{\smallskip}
C&$3$&$-37.7$&0.43&1.10&0.42&0.82&0.82\\
\noalign{\smallskip}
&$4$&$-75.6$&0.17&1.02&0.17&0.88&0.88\\
\noalign{\smallskip}
&$5$&$-110.3$&0.15&1.02&0.15&0.88&0.88\\
\toprule[0.8pt] \noalign{\smallskip}
\end{tabular}
\end{table}

If we increase the ratio $\frac{m_Q}{m_b}$, the sizes of the
subclusters are convergent in each configuration, see
$\langle\mathbf{r}^2\rangle^{\frac{1}{2}}$ and
$\langle\mathbf{R}^2\rangle^{\frac{1}{2}}$ in Table~\ref{variation}.
However, the distance between two heavy quarks
$\langle\mathbf{r}_{QQ}^2\rangle^{\frac{1}{2}}$ decreases gradually
till they shrink into a tiny and compact core eventually. The two
subclusters overlap completely. The three configurations will
degenerate into a single one. Its size can be approximately
described by either the $\langle\mathbf{r}^2\rangle^{\frac{1}{2}}$
or $\langle\mathbf{R}^2\rangle^{\frac{1}{2}}$. The $QQ$-core
contributes to the vast majority of the binding energy of the doubly
heavy tetraquark states. The light quarks $\bar{u}$ and $\bar{d}$
move around the $QQ$-core. Their relative distance
$\langle\mathbf{r}_{\bar{u}\bar{d}}^2\rangle^{\frac{1}{2}}$ is about
1 fm. In summary, the doubly heavy tetraquark states look like a
helium-like QCD-atom in the limit of a large heavy quark mass.

\section{Other partner states of the $T^+_{cc}$ }

\subsection{Isospin antisymmetric states $T^0_{bc}$}

The state $T^0_{bc}$ with $00^+$ can form a shallow bound state with
the $E_b$ of several or a dozen MeV in the four cases, in which the
dominant component is the channel $\bar{B}D$. Our conclusion is very
close to that of other model
calculations~\cite{tcc-lee,tbc-vijande}. In the cases (a) and (b), the
overlapping between two subclusters is very obvious so that the
state looks like a compact state because of the strong attraction
coming from the $V^{oge}$ and $V^{\sigma}$. Turning off the meson
exchange interactions, the $T^0_{bc}$ state with $00^+$ becomes a
loosely bound state in the cases (c) and (d) while a hydrogen
molecule-like state does not appear.

The nonidentity of the $b$ and $c$ quark in the $T^0_{bc}$ state
with $01^+$ enlarges the Hilbert space comparing with the $T^-_{bb}$
case with $01^+$. Now we have to consider three coupling channels.

From Table~\ref{partners}, the $01^+$ state has a binding energy of
several MeV in the four cases, which is slightly lower than the
lattice QCD results in the range of 20-40 MeV below the $\bar{B}^*D$
threshold~\cite{tbclattice}. The dominant component of the state is
the $\bar{B}^*D$ channel, which is supported by other model
predictions~\cite{tcc-lee,tbc-vijande}. The $T^0_{bc}$ state with
$01^+$ is a shallower bound state than the $T^-_{bb}$ with $01^+$
because of the lighter charm quark mass and thus larger kinetic
energy. Moreover, the overlapping between the two subclusters,
binding energy and the delocalization effect of the light quarks
become weaker. However, the physical picture such as the emergence
of the compact state, hydrogen molecule-like state or deuteron-like
state in four cases in the $T^0_{bc}$ system resembles that of the
$T^-_{bb}$ state with $01^+$.

The mass of the $T^0_{bc}$ state with $02^+$ is about 12 MeV lower
than the $\bar{B}^*\bar{D}^*$ threshold due to its small kinetic
energy $E_k$ in the model, see the cases (a) and (c) in
Table~\ref{partners}. This state can form a hydrogen-like state
similar to the $T^-_{bb}$ with $01^+$ due to the delocalization of
the light quarks induced by the color screening effect in the
confinement. In the cases (b) and (d), the state has a binding
energy of 3.6 MeV relative to the $\bar{B}^*\bar{D}^*$ threshold,
which is a deuteron-like state because the two subclusters are
separated well apart. The $T^0_{bc}$ state with $02^+$ is not stable
although it is below the $\bar{B}^*D^*$ threshold because it
can decay into the modes $\bar{B}D$, $\bar{B}D\pi$,
$\bar{B}^*D\gamma$, $\bar{B}D^*\gamma$, $\bar{B}D\gamma\gamma$ etc.
However, the state should be very narrow.

\subsection{V-spin antisymmetric states $T^-_{bbs}$ and $T^0_{bcs}$}

The corresponding SU(2) groups of the $I$-spin, and the so-called
$V$-spin and $U$-spin are three subgroups of the flavor SU(3) group.
Therefore, the $V$-spin antisymmetric $T^-_{bbs}$ with
$\frac{1}{2}1^+$ and the state $T^-_{bb}$ with $01^+$ should share
the same symmetry in their wave functions so that their behaviors
should be analogous from the perspective of quark models. Similar
arguments hold for the $V$-spin antisymmetric state $T^0_{bcs}$ and
the state $T^0_{bc}$ with $I=0$.

From Table~\ref{partners}, the $V$-spin antisymmetric state
$T^-_{bbs}$ with $\frac{1}{2}1^+$ is a shallow bound state with a
binding energy about 10 MeV relative to the threshold
$\bar{B}\bar{B}^*_s$, where the attraction mainly comes from the
$V^{\sigma}$ in the cases (a) and (b). Our results agree with those
in Ref. \cite{tcc-lee} but are less than the latest lattice QCD
results about 80 MeV ~\cite{tbbslattice}. Similar to the state
$T^-_{bb}$ with $01^+$, the state $T^-_{bbs}$ with $\frac{1}{2}1^+$
can also form a compact state in the cases (a) and (b), which should
therefore be a compound of color singlet and hidden color states.
The lattice result indicated that the meson-meson percentage is
about 84\% while the hidden color percentage is about
16\%~\cite{tbbslattice}. In the case (c), the state $T^-_{bbs}$ is a
hydrogen molecule-like bound state because of the delocalization of
the light quark $\bar{u}$ and $\bar{s}$ induced by the color screen
effect. In the case (d), the state $T^-_{bbs}$ forms a deuteron-like
state because of removing the meson exchange interaction and color
screening effect from the model.

Both of the V-spin antisymmetric states $T^0_{bcs}$ with
$\frac{1}{2}0^+$ and $\frac{1}{2}1^+$ appear a bound state in the
model, which is qualitatively consistent with the conclusions in
Refs.~\cite{tcc-lee,tbclattice,tbcuslattice}. In the cases (a) and (b),
the two states are shallow bound states with a binding energy of
several MeV because of the $V^{\sigma}$. The two subclusters have a
slight or even no overlapping. In the cases (c) and (d), the two
states are very loosely bound without the meson exchange
interactions. The two states are deuteron-like states because the
subclusters are completely separated from each other. The hydrogen
molecule-like configuration appearing in the state $T^-_{bbs}$ with
$01^+$ vanishes in the $T^0_{bcs}$ states with $\frac{1}{2}0^+$ and
$\frac{1}{2}1^+$.

The state $T^0_{bcs}$ with $\frac{1}{2}2^+$ is lower than the
corresponding threshold $\bar{B}_s^*\bar{D}^*$ in the model. Similar
to the state $T^0_{bc}$ with $02^+$, the state $T^0_{bcs}$ with
$\frac{1}{2}2^+$ is also a hydrogen molecule-like bound state
because of the delocalization of light quarks in the cases (a) and
(c). Removing the color screen effect from the model, the state
$T^0_{bcs}$ with $\frac{1}{2}2^+$ becomes a deuteron-like bound
state in the cases (b) and (d). The state is not stable and can
decay into $\bar{B}_sD$, $\bar{B}_sD\pi$, $\bar{B}_sD^*\gamma$,
$\bar{B}_s^*D\gamma$, $\bar{B}_sD\gamma\gamma$.

\subsection{Other unstable states}

All of the isospin symmetric states $T^+_{cc}$, $T^-_{bb}$,
$T^0_{bc}$, $T^{++}_{ccss}$, $T^{0}_{bbss}$ and $T^{+}_{bcss}$
cannot form bound states because the interactions can not provide
enough attraction in the model. Other model studies on the states
$T^0_{bc}$ and $T^{++}_{ccss}$ also suggest that the isospin
symmetric states are unbound and
unstable~\cite{tcc-lee,tbc-vijande,tccss-yang,tccss-ping}. The
lattice QCD investigations on the isospin symmetric states $T^+_{cc}$,
$T^-_{bb}$, $T^0_{bc}$, $T^{++}_{ccss}$, $T^{0}_{bbss}$ and
$T^{+}_{bcss}$ indicated that no clear signal of any level below
their respective thresholds can be found~\cite{tbblattice3}. Similarly,
the V-spin symmetric states $T^+_{ccs}$, $T^-_{bbs}$ and $T^0_{bcs}$
can not form stable bound states in the model. The V-spin
antisymmetric state $T^+_{ccs}$ with $\frac{1}{2}1^+$, the strange
partner of the state $T^+_{cc}$ with $01^+$, could be a stable bound
state in some theoretical frameworks~\cite{tcc-lee,tccs,tccsbseq}.
However, the state is not stable in the present model. The situation
may change if the mixing of $S$-$D$ wave is taken into account in
the model, which is left for the future work.

\section{Magnetic moments and axial charges}

The magnetic moments of hadrons encode useful information about the
distributions of the charge and magnetization inside the hadrons,
which help us to understand their geometric configurations. Ignoring
the contributions from the quark orbital angular momentum, the
operator for the magnetic moment of the doubly heavy tetraquark
system is given simply by
\begin{eqnarray}
\hat{\mu}_{m}=\sum_{i=1}^4\frac{\hat{Q}_i}{2m_i}\hat{\sigma}_i^z,\nonumber
\end{eqnarray}
where $\hat{Q}_i$ is the electric charge operator of the $i$-th
quark and the $\sigma^z_i$ is the $z$-component of Pauli matrix. We
can obtain the magnetic moments of the doubly heavy tetraquark
states below their corresponding threshold by directly calculating
the matrix element
\begin{eqnarray}
\mu_{m}=\langle \Phi_{IJ}|\hat{\mu}_{m}|\Phi_{IJ}\rangle,\nonumber
\end{eqnarray}
where $\Phi_{IJ}$ is the eigenvector of those states.

From Table~\ref{magnetic}, the spin and magnetic momentums of the
states $T^0_{bc}$ with $00^+$ and $T^{+}_{bcs}$ with
$\frac{1}{2}0^+$ vanish. The magnetic momentums of the other states
depend on their spatial configurations except for the two high spin states.
The magnetic momentum of the $T^+_{cc}$ state  with $01^+$ is about 0.18
$\mu_N$ and 0.13 $\mu_N$ in the cases (a) and (b), respectively. The magnetic
momentum of the $T^+_{cc}$ state with $01^+$ was also studied using the light-cone
QCD sum rule formalism ~\cite{tcc-magmom1}. Its value was roughly
$0.66$ $\mu_N$ and $0.43$ $\mu_N$ for the compact
diquark-antidiquark and molecule pictures, respectively in Ref.
~\cite{tcc-magmom1}.

The axial charge $g_A$ is an important quantity for the
understanding of both the electroweak and strong interactions. The
nonrelativistic leading order axial charge operator for a point-like
Dirac constituent quark is given by the Gamov-Teller operator
$\hat{\sigma}^z\hat{\tau}^z$~\cite{axial-charge}, where $\tau^z$ is
the isospin operator. Then the axial charge operator for the doubly
heavy tetraquark states is given by
\begin{eqnarray}
\hat{g}_{A}=\sum_{i=1}^4\hat{\tau}_i^z\hat{\sigma}_i^z. \nonumber
\end{eqnarray}
In this way, we can achieve the axial charge $g_A$ of the states by
directly calculating the matrix element
\begin{eqnarray}
g_{A}=\langle \Phi_{IJ}|\hat{g}_{A}|\Phi_{IJ}\rangle,\nonumber
\end{eqnarray}
which are presented in Table~\ref{magnetic}. The axial charges of
the states $T^+_{cc}$, $T^-_{bb}$ and $T^0_{bc}$ are zero because
their isospin magnetic components are zero while that of $T^0_{bcs}$
with $\frac{1}{2}0^+$ is zero because its spin magnetic component is
zero. The axial charges of the other states also depend on their
spatial configurations but do not change dramatically in the four
cases.
\begin{table}[h]
\caption{Magnetic moment $\mu_m$ unit in $\mu_N$ and axial charge $g_A$ unit in $g_V$ in the four cases.}\label{magnetic}
\begin{tabular}{cccccccccccccccccc}
\toprule[0.8pt]
\noalign{\smallskip}
State&$T^+_{cc}$&$T^-_{bb}$&$T^0_{bc}$&&&$T^-_{bbs}$&$T^0_{bcs}$&&\\
\noalign{\smallskip}
$IJ^P$&~~$01^+$~~&$01^+$&~~$00^+$~~&$01^+$&~~$02^+$~~&$\frac{1}{2}1^+$&~~$\frac{1}{2}0^+$~~&$\frac{1}{2}1^+$&~~$\frac{1}{2}2^+$\\
\noalign{\smallskip}
\toprule[0.8pt]
\noalign{\smallskip}
$\mu^a_m$&0.18&0.64&0.00&0.66&0.79&1.37&0.00&0.98&~1.29\\
\noalign{\smallskip}
$\mu^b_m$&0.13&0.49&0.00&0.59&0.79&1.29&0.00&0.94&~1.29\\
\noalign{\smallskip}
$\mu^c_m$&-&0.97&0.00&0.72&0.79&1.40&0.00&0.95&~1.29\\
\noalign{\smallskip}
$\mu^d_m$&-&0.98&0.00&0.67&0.79&1.37&0.00&1.13&~1.29\\
\noalign{\smallskip}
\toprule[0.8pt]
\noalign{\smallskip}
$g^a_A$&0.00&0.00&0.00&0.00&0.00&0.81&0.00&1.13&~2.18\\
\noalign{\smallskip}
$g^b_A$&0.00&0.00&0.00&0.00&0.00&0.74&0.00&1.09&~2.13\\
\noalign{\smallskip}
$g^c_A$&0.00&0.00&0.00&0.00&0.00&0.83&0.00&1.09&~2.15\\
\noalign{\smallskip}
$g^d_A$&0.00&0.00&0.00&0.00&0.00&0.79&0.00&1.03&~2.08\\
\noalign{\smallskip}
\toprule[0.8pt]
\noalign{\smallskip}
\end{tabular}
\end{table}

\section{summary}

In the present work, we have performed a systematical investigation
of the doubly heavy tetraquark states with the molecule
configuration within the framework of the nonrelativistic quark
model with the help of the Gaussian expansion method. The model
includes the color screening confinement potential, meson-exchange
interactions and one-gluon-exchange interactions. Besides the
tetraquark spectrum and their spatial configurations, we have also
calculated the magnetic moments and axial charges of the statable
doubly heavy tetraquark states.

We discuss various dynamical effects in the formation of the stable
bound states against the strong interactions extensively. We
decompose the attractions from various sources and illustrate the
very delicate competition between the kinetic energy and attractive
potentials in the formation of three kinds of different bound
states: the compact, deuteron-like or hydrogen molecule-like states.

The dominant component of the recently discovered $T^+_{cc}$ state
by the LHCb Collaboration is the $DD^*$ component, which can not
form a bound state alone in the model. The coupled channel effect
between the $[DD^*]_-$ and $[D^*D^*]_-$ channels plays a critical
role in the formation of the $T^+_{cc}$ under the assumption of
meson-meson picture. The long-range $\pi$ and intermediate-range
$\sigma$ exchange interactions also play a pivotal role. Without the
meson exchange force, the $T^+_{cc}$ states does not exist. With the
model parameters extracted from the ordinary meson spectrum and
without introducing any new parameters in the present calculation,
we extracted the binding energy of the $T^+_{cc}$ to be 0.34 MeV,
which agrees with LHCb's measurement very well. With a huge size
around 4.32 fm, the $T^+_{cc}$ state is a loosely bound
deuteron-like state.

There is a broad theoretical consensus that the tendency to form
doubly heavy tetraquark bound states is proportional to the mass
ratio $\frac{m_Q}{m_q}$. In the limit of large masses of the heavy
quarks the corresponding ground state should be deeply bound. Such a
qualitative picture is strengthened by our numerical results. The
existence of the shallow bound state $T^+_{cc}$ implies that there
should exist many stable doubly heavy tetraquark states.

Our investigations indicate that the $I$-spin antisymmetric states
$T^-_{bb}$ with $01^+$, $T^0_{bc}$ with $00^+$ and $01^+$, the
$V$-spin antisymmetric states $T^-_{bbs}$ with $\frac{1}{2}1^+$,
$T^0_{bcs}$ with $\frac{1}{2}0^+$ and $\frac{1}{2}1^+$ can form a
compact, hydrogen molecule-like, or deuteron-like bound state
depending on different binding dynamics. The compact spatial size of
the $T^-_{bb}$ may require the introduction of the hidden-color
configuration from the very beginning, which is the topic of our
future work.

The high-spin states $T^0_{bc}$ with $02^+$ and $T^0_{bcs}$ with
$\frac{1}{2}2^+$ can decay into $D$-wave $\bar{B}D$ and $\bar{B}_sD$
through the strong interactions although they are below the
thresholds $\bar{B}^*D^*$ and $\bar{B}^*_sD^*$, respectively. The
$I$-spin or $V$-spin symmetric states, $T^+_{cc}$, $T^0_{bc}$,
$T^-_{bb}$, $T^0_{bcs}$, $T^-_{bbs}$, $T^{++}_{ccss}$,
$T^{+}_{bcss}$ and $T^{0}_{bbss}$, are unbound in the model
prediction. The state $T^+_{ccs}$ is also not bound in the model no
matter its $V$-spin is symmetric or antisymmetric, which may be due
to the omission of the $S$-$D$ wave mixing in the present work.

The discovery of the $T^+_{cc}$ state opened a new window for hadron
physics. More theoretical and experimental efforts are called for in
order to understand its underlying structure and nonperturbative QCD
dynamics in this region. We sincerely hope that some of the doubly
heavy tetraquark candidates may be searched for at LHCb and
BelleII in the near future.

\acknowledgments {One of the authors C. Deng thanks Prof. J.L. Ping
for helpful discussions. This research is partly supported by the
National Science Foundation of China under Contracts No. 11975033
and No. 12070131001, Chongqing Natural Science Foundation under
Project No. cstc2019jcyj-msxmX0409 and Fundamental Research Funds
for the Central Universities under Contracts No. SWU118111.}

\end{document}